# BIG IOT AND SOCIAL NETWORKING DATA FOR SMART CITIES:

*Algorithmic improvements on Big Data Analysis in the context of RADICAL city applications*

Evangelos Psomakelis[1,2], Fotis Aisopos[1], Antonios Litke[1], Konstantinos Tserpes[2,1], Magdalini Kardara[1] and Pablo Martínez Campo[3]

[1]*Distributed Knowledge and Media Systems Group, National Technical University of Athens, Zografou Campus, Athens, Greece*
[2]*Informatics and Telematics Dept, Harokopio University of Athens, Greece*
[3]*Communications Engineering department, University of Cantabria, Santander, Spain*
*{fotais, litke, nkardara, tserpes, vpsomak}@mail.ntua.gr, pmartinez@tlmat.unican.es*

Keywords: Internet of Things, Social Networking, Big Data Aggregation and Analysis, Smart City applications, Sentiment Analysis, Machine Learning

Abstract: In this paper we present a SOA (Service Oriented Architecture)-based platform, enabling the retrieval and analysis of big datasets stemming from social networking (SN) sites and Internet of Things (IoT) devices, collected by smart city applications and socially-aware data aggregation services. A large set of city applications in the areas of Participating Urbanism, Augmented Reality and Sound-Mapping throughout participating cities is being applied, resulting into produced sets of millions of user-generated events and online SN reports fed into the RADICAL platform. Moreover, we study the application of data analytics such as sentiment analysis to the combined IoT and SN data saved into an SQL database, further investigating algorithmic and configurations to minimize delays in dataset processing and results retrieval.

## 1 INTRODUCTION

Modern cities are increasingly turning towards ICT technology for confronting pressures associated with demographic changes, urbanization, climate change (Romero Lankao, 2008) and globalization. Therefore, most cities have undertaken significant investments during the last decade in ICT infrastructure including computers, broadband connectivity and recently sensing infrastructures. These infrastructures have empowered a number of innovative services in areas such as participatory sensing, urban logistics and ambient assisted living. Such services have been extensively deployed in several cities, thereby demonstrating the potential benefits of ICT infrastructures for businesses and the citizens themselves. During the last few years we have also witnessed an explosion of sensor deployments and social networking services, along with the emergence of social networking (Conti et al., 2011) and internet-of-things technologies (Perera et al., 2013; Sundmaeker et al., 2010) Social and sensor networks can be combined in order to offer a variety of added-value services for smart cities, as has already been demonstrated by various early internet-of-things applications (such as WikiCity(Calabrese et al., 2007), CitySense(Murty et al., 2007), GoogleLatitude(Page and Kobsa, 2010)), as well as applications combining social and sensor networks (as for example provided by (Breslin and Decker, 2007; Breslin et al., 2009) and (Miluzzo et al., 2007). Recently, the benefits of social networking and internet-of-things deployments for smart cities have also been demonstrated in the context of a range of EC co-funded projects (Hernández-Muñoz et al., 2011; Sanchez, 2010).

Current Smart City Data Analysis implies a wide set of activities aiming to turn into actionable data the outcome of complex analytics processes. This analysis comprises among others: i) analysis of thousands of traffic, pollution, weather, waste, energy and event sensory data to provide better services to the citizens, ii) event and incident analysis using near real-time data collected by citizens and devices sensors, iii) turning social media data related to city issues into event and sentiment analysis , and many others. Combining data from physical (sensors/devices) and social sources (social networks) can give more complete, complementary data and contributes to better analysis and insights. In overall, smart cities are complex social systems and large scale data analytics can contribute into their sustainability, efficient operation and welfare of the citizens.

Motivated by the modern challenges in smart cities, the RADICAL approach (RADICAL, 2016) opens new horizons in the development, deployment and operation of interoperable social networking and Internet of Things services in smart cities, notably services that could be flexibly and successfully customized and replicated across multiple cities. Its main goal is to provide the means for cities and SMEs to rapidly develop, deploy, replicate, and evaluate a diverse set of sustainable ICT services that leverage established IoT and SN infrastructures. Application services deployed and pilotedinvolve: i) Cycling Safety Improvement, ii) Products Carbon Footprint Management, iii) Object-driven Data Journalism, iv) Participatory Urbanism, v) Augmented Reality, vi) Eco- consciousness, vii) Sound map of a city and viii) City-R-Us: a crowdsourcing app for collecting movement information using citizens smartphones.

The RADICAL platform is an open platform having as added value the capability to easily replicate the services in other smart cities, the ability to co-design services with the involvement of cities' Living Labs, and the use of added value services that deal with the application development, the sustainability analysis and the governance of the services.

The RADICAL approach emphasizes on the sustainability of the services deployed, targeting both environmental sustainability and business viability. Relevant indicators (e.g., CO2 emissions, Citizens Satisfaction) are established and monitored as part of the platform evaluation.End users (citizens) in modern smart cities are increasingly looking for media-rich services offered under different space-, context-, and situational conditions. The active participation and interaction of citizens can be a key enabler for successful and sustainable service deployments in future cities. Social networks hold the promise to boost such participation and interaction, thereby boosting participatory connected governance within the cities. However, in order to enable smart cities get insight information on how citizens think, act and talk about their city it is important to understand their opinion and sentiment polarity on issues related to their city context. This is where sentiment analysis can play a significant role. As social media data bring in significant Big Data challenges (especially for unstructured data streams) it will be important to find effective ways to analyse sentimentally those data for extracting value information and within specific time windows.

This paper has the following contributions:
- Innovative smart city infrastructure for uniform social and IoT big data aggregation and combination.
- Comperative study over Sentiment Analysis techniques efficiency, to reduce record, retrieval, update and processing time.
- A novel technique for n-grams storage and frequency representation in the context of big data Sentiment Analysis.

The rest of the paper is structured as follows: Section 2 gives an overview of related and similar works that can be found in the international literature and in projects funded by the European Commission. Section 3 presents the RADICAL architecture and approach. Section 4 presents details about the Sentiment Analysis problem and related experiments, while in section 5 we provide the future work to be planned in the context of RADICAL and the conclusions we have come into.

## 2 RELATED WORK

Recently, various analytical services such as sentiment analysis found their way into Internet of Things (IoT) applications. With the devices that are able to convey human messages over the internet meeting an exponential growth, the challenge now revolves around big data issues. Traditional approaches do not cope with the requirements posed from applications for analytics in e.g. high velocity rates or data volumes. As a result, the integration of IoT with social sensor data put common tasks like feature extraction, algorithm training or model updating to the test.

Most of the algorithms are memory-resident and assume a small data size (He et al., 2010) and once this threshold is exceeded, the algorithms' accuracy and performance degrades to the point they are useless. Therefore even if we focused solely on volume challenges, it is intuitively expected that the accuracy of the supervised algorithms will be affected. An attempt from (Liu et al., 2013) to use Naïve Bayes in an increasingly large data volume, showed that a rapid fall of the algorithms accuracy is followed by a continuous, smooth increase asymptotically tending from the lower end to the baseline (best accuracy under normal data load).

Rather than testing the algorithm's limitations, most of the other approaches are focusing on implementing parallel and distributed versions of the algorithms such as (He et al., 2010; Read et al., 2015). In fact most of them rely on the Map-Reduce framework so as to achieve high throughput classification (Amati et al., 2014; Sakaki et al., 2013; Wang et al., 2012; Zhao et al., 2012) whereas a number of toolkits have been presented with implementations of distributed or parallel versions of machine learning algorithms such as ("Apache Mahout: Scalable machine learning and data mining," n.d., "MEKA: A Multi-label Extension to WEKA," n.d.; Bifet et al., 2010). While these solutions put most of the emphasis in the model and the optimization of the classification task in terms of accuracy and throughput, there is a rather small body of research dealing with the problem of feature extraction in high pace streams. The standard solution that is considered is the use of a sliding window and the application of standard feature extraction techniques in this small set. In cases where the stream's distribution is variable, a sliding window kappa-based measure has been proposed (Bifet and Frank, 2010).

As reported in (Strohbach et al., 2015), another domain of intense research in the area of scalable analytics is for an architecture that combines both batch and stream processing over social and IoT data while at the same time considering a single model for different types of documents (e.g. tweets Vs blogposts). Sentiment analysis is a typical task that requires batch modeling in order to generate the golden standards for each of the classes. This process is also the most computationally intense, as the classification task itself is usually a CPU bound task (i.e. run the classification function). In a data streaming scenario the golden standards must be updated in a batch mode, whereas the feature extraction and classification must take place in real time.

Perhaps the most prominent example of such an architecture is the Lambda Architecture (Marz and Warren, 2015) pattern which solves the problem of computing arbitrary functions on arbitrary data in realtime by combining a batch layer for processing large scale historical data and a streaming layer for processing items being retrieved in real time from an input queue or analytics in e.g. high velocity rates or data volumes. As a result, the integration of IoT with social sensor data put common tasks like feature extraction, algorithm training or model updating to the test.Most of the algorithms are memory-resident and assume a small data size (He et al., 2010) and once this threshold is exceeded, the algorithms' accuracy and performance degrades to the point they are useless. Therefore even if we focused solely on volume challenges, it is intuitively expected that the accuracy of the supervised algorithms will be affected. An attempt from Liu et al (Liu et al., 2013) to use Naïve Bayes in an increasingly large data volume, showed that a rapid fall of the algorithms accuracy is followed by a continuous, smooth increase asymptotically tending from the lower end to the baseline (best accuracy under normal data load).

## 3. THE RADICAL APPROACH

The RADICAL platform integrates components and tools from (SocIoS, 2013) and (SmartSantander, 2013) projects, in order to support innovative smart city services, leveraging information stemming from Social Networks (SN) and Internet of Things devices. Using the aforementioned tools, it can collect, combine, analyze, process, visualize and provide uniform access to big datasets of Social Network content (e.g. tweets) and Internet of Things information (e.g. sensor measurements or citizen smartphone reports).

The architecture of the RADICAL platform is depicted in Figure 1. As can be observed, all IoT data are pushed into the platform through the respective Application Programming Interfaces (IoT API and Repository API) and are forwarded to the RADICAL Repository, comprised by a MySQL database, formed based on the RADICAL Object Model. The device-related data, as dictated by this object model, are saved in the form of Observations and Measurements. Observations correspond to general IoT events reported (e.g. a sensor report or bicycle "check-in" event), while Measurements to more specific metrics included in an Observation (e.g. Ozone measurements (mpcc) or bicycle current speed (km/h)). On the other hand, SN data are accessed in real time from the underlying SN adaptors, by communicating with the respective Networks' APIs. In cases of Social Networks like Foursquare that provide plain venues and statistics, the adaptor-like data structures do not make sense, thus relevant Social Enablers are used to retrieve venue-related information data.

On top of the main platform, RADICAL delivers a set of tools (Application Management layer) that allow end users to make better use of the RADICAL platform, such as configuring the registered IoT devices or extracting general activity statistics, through the RADICAL Configuration API. Lastly, the RADICAL Data API allows smart city services to access the different sources of information (social networks, IoT infrastructures, city applications), combine data and perform data analysis by using the appropriate platform tools.

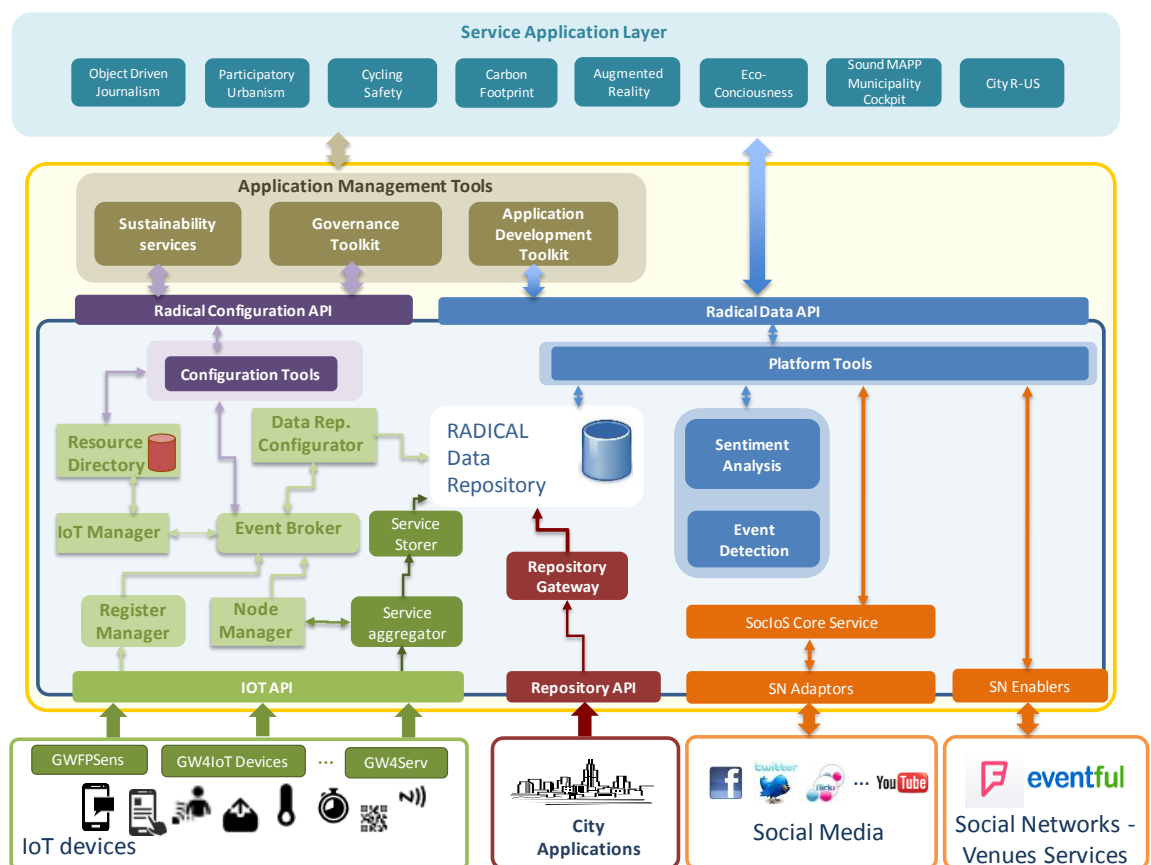

Figure 1: RADICAL Platform Architecture

As can be seen in the Service Application Layer, in the context of RADICAL a wide range of Smart City services of various scopes has been developed:

- **Citizen journalism and Participatory Urbanism**: Those two interrelated services allow citizens reporting events of interest in the city, by posting images, text and metadata through their smartphones.
- **Cycling Safety**: Cyclists, acting as human sensors can report the situation in the city streets through their smartphones.
- **Monitoring the carbon footprint of products, people and services**: By using a range of sensors, the CO2 emissions in specific places in a city may be monitored.
- **Augmented Reality in Points of Interest (POI)**: Tourists use their smartphones to identify and receive information about points of interest in a city.
- **Propagation of eco-consciousness:** Leverages on the viral effect in the propagation of information in the social networks as well as the recycling policy of a city, through monitoring and reporting relevant actions on citizens' smartphones.
- **Social-Orientated Urban Noise Decibel Measurement Application**: Noise sensors are employed throughout the city and citizens are able to report and comment noise-related information through SNs under a hashtag.
- **City Reporting application for the use of Urban Services**: This service gathers sensory data along with SN check-ins in city venues, to construct a traffic map throughout the city, leveraging the process load of any centralized decision making process.

The aforementioned services are piloted in six European participating cities: Aarhus, Athens, Genoa, Issy les Moulineaux, Santander and the region of Cantabria. Figure 2 illustrates a screenshot example of the RADICAL Cities' Dashboard, where general statistics on device registration and activity for a service throughout different cities in a specific time period is provided. In overall, during the last pilot iteration, RADICAL Repository had captured a total of **5.636** active IoT devices sending **728.253** Observations and **5.461.776** Measurements.

Most of the services above depend on the aggregation of those IoTdata with social data stemming from online Social Network sites. E.g. in the Participatory Urbanism service, citizens' reports sent through smartphones and saved in RADICAL Repository are combined with relevant tweets (under a city service hashtag), as well as POI information that can be collected from similar SNs.

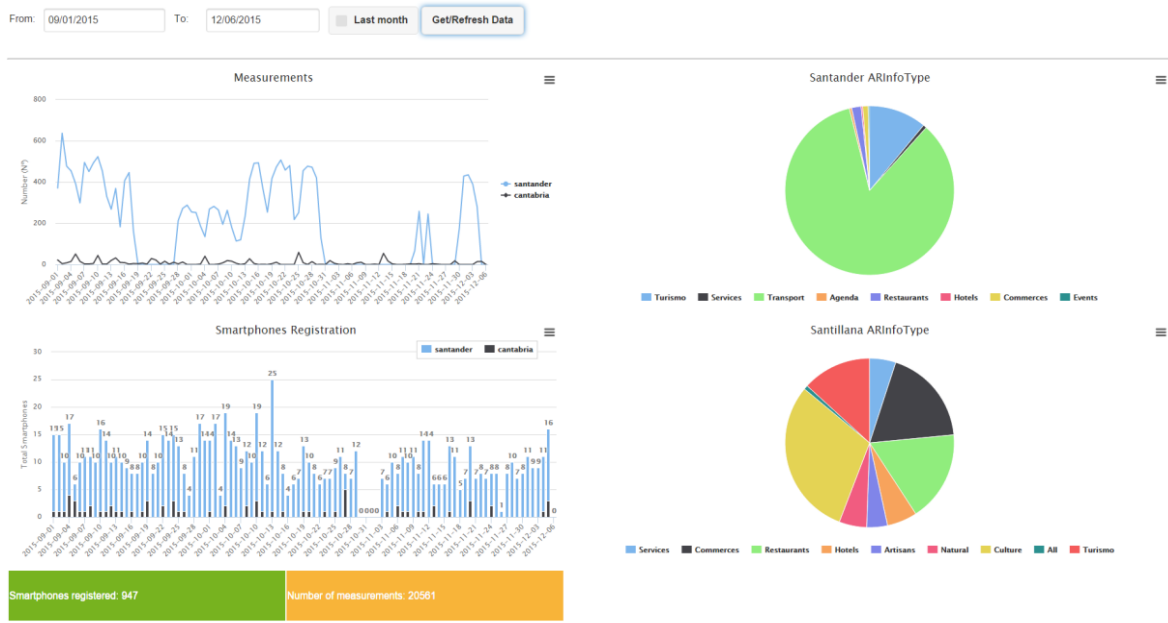

Figure 2: RADICAL Cities Dashboard presents smartphone registrations and measurements for the AR service in the cities of Santander and Cantabria over a period

Thus, given the size of the datasets acquired by smart city services, along with the rich social media content that can be retrieved through the RADICAL platform adaptors, big data aggregation and analysis challenges arise. Data Analysis tools are the ones that further process the data in order to provide meaningful results to the end user, i.e. Event Detection or Sentiment Analysis.

When it comes to Big Data, as in the RADICAL case, where millions of user-reported events are aggregated along with millions of SN posts and an extraction of general results is required, the challenge accrued is two-fold: First, the tool must ensure the accuracy of the analysis, in the sense that data classification is correct to a certain and satisfactory extent, and second, processing time must be kept under certain limits, so that results retrieval process delay is tolerable by end-users. Moreover, it is apparent in such analysis that a trade-off between effectiveness and efficiency exists. The latter is a most crucial issue in Big Data analysis and apart from the policy followed in data querying (e.g. for queries preformed in an SQL database), it is also related to the algorithmic techniques employed foranalyzing those datasets.

In the context of this work, we focus on the Sentiment Analysis on the big IoT and SN related datasets of RADICAL, as this was the most popular functionality among participating cities and almost all of the RADICAL Smart City services presented above make use of it. The goal of the Sentiment Analysis service is to extract sentiment expressive patterns from user-generated content in social networks or IoT-originated text posts. The service comes to the aid of the RADICAL city administrators, helping them to categorize polarized posts, meaning sentimentally charged text, e.g. analyse citizens' posts to separate subjective from objective opinions or count the overall positive and negative feedback, concerning a specific topic or event in the city.

# 4. SENTIMENT ANALYSIS EXPERIMENTS AND PERFORMANCE IMPROVEMENT

## 4.1 Introduction

The term Sentiment Analysis refers to an automatic classification problem. Its techniques are trying to distinguish between sentences of natural language conveying positive (e.g. happiness, pride, joy), negative (e.g. anger, sadness, jealously) or even neutral (no sentiment texts like statements, news, reports)emotion (called sentiment for our purposes) (Pang et al., 2002).

A human being is capable of understanding a great variety of emotions from textual data. This process of understanding is based on complicated learning procedures that we all go through while using our language as a means of communication, be it actively or passively. It requires imagination and subjectivity in order to fully understand the meaning and hidden connections of each word in a sentence, two things that machines lack.

The most common practice is to extract numerical features out of the natural language (Godbole et al., 2007). This process translates this complex means of communication into something the machine can process.

## 4.2 Natural Language Processing

In order to process the natural language data, the computer has to take some pre-processing actions. These actions include the cleansing of irrelevant, erroneous or redundant data and the transformation of the remaining data in a form more easily processed.

Cleansing the data has become a subjective task, depending on the purposes of each researcher and the chosen machine learning algorithms. The transformation of the sentences in another form now is clearly studied and each approach has some advantages and disadvantages. This paper will detail three approaches, two widely used and one that had some success in improving the accuracy of the algorithms: the bag of word, N-Gramsand N-Gram Graphs(Aisopos et al., 2012; Fan and Khademi, 2014; Giannakopoulos et al., 2008; Pang and Lee, 2008).

The bag of words approach is perhaps the most simple and common one. It regards each sentence as a set of words, disregarding their grammatical connections and neighbouring relations. It splits each sentence based on the space character (in most languages) and then forms a set of unrelated words (a bag of words as it is commonly called). Then each word in this bag can be disregarded or rated by a numerical value, in order to create a set of numbers instead of words.

The N-Grams are a bit more complex. They also form a bag of words but now each sentence is split into pseudo-words of equal length. A sliding window of N characters is rolling on the sentence creating this bag of pseudo-words. For example if N=3 the sentence "This is a nice weather we have today!" will be split in the bag {'Thi', 'his', 'is ', 's i', ' is', 'is ', 's a', ' a ', 'a n', ' ni', 'nic', 'ice', 'ce ', 'e w', ' we', 'wea', 'eat', 'ath', 'the', 'her', 'er ', 'r w', ' we', 'we ', 'e h', ' ha', 'hav', 'ave', 've ', 'e t', ' to', 'tod', 'oda', 'day', 'ay!'}.

This technique takes into regard the direct neighbouring relations by creating a continuous stream of words, it still ignores the indirect relations between words and even the relations between the produced N-Grams. Of course it is impossible to have a predefined set of numerical ratings for each one of these pseudo-words because each sentence and each N number (which is defined arbitrarily by the researcher) produces a different set of pseudo-words(Psomakelis et al., 2014). So machine learning is commonly used to replace these words with numerical values and create sets of numbers which can be aggregated to sentence level.

An improvement on that approach aims to take into consideration the neighbouring relations between the produced N-Grams. This approach is called N-Gram Graphs and its main concept is to create a graph connecting each N-Gram with its neighbours in the original sentence. So each node in this graph is an N-Gram and each edge is a neighbouring relation(Giannakopoulos et al., 2008). This approach gives a variety of new informationto the researchers and to the machine learning algorithms, including information about the context of words, making it a clear improvement of the simple N-Grams(Aisopos et al., 2012). The only drawbacks are the complexity it adds to the process and the difficulties of storing, accessing and updating a graph of textual data.

## 4.3 Dataset Improvements

At the core of sentiment analysis is its dataset. We are gathering and employing bigger and bigger datasets in order to better train the algorithms to distinguish what is positive and what is negative. Classic storage techniques are proving more and more cumbersome for large datasets. ArrayLists and most Collections are adding a big overhead to the data so they are not only enlarging the space requirements for its storage but they are also delaying the analysis process. So new techniques for data storage and retrieval are needed, techniques that will enable us to store even bigger datasets and access them with even smaller delays.

The most commonly used such technique is the Hash List(Fan and Khademi, 2014), which first hashes the data in a certain, predefined amount of buckets and then creates a List in each bucket to resolve any collisions. This method's performance is heavily dependent on the quality of the hash function and its ability to equally split the data into the buckets. The target is to have as small lists as possible. That is the case because finding the right bucket for a certain piece of data is done in O(1) time but looking through the List in that bucket for the correct spot to store the piece of data is done in O(n) time where n is the number of data pieces in the List.

Moreover, in Java which is the programming language that we are using, each List is an object containing one object for each data piece. All these objects create an overhead that is not to be ignored. In detail the estimated size that a hash list will occupy is calculated as:

$$12 + ((B - E) * 12) + (E * 4) + (U * (N * 2 + 72))$$

Equation 1: Size estimation of Hash List where N=NGram Length, U=Unique NGrams, B=Bucket Size, E=Empty Buckets.

The worst case for storage but best for access time is when almost each data piece has its own bucket. In this case, for N=5, S=11881376, U=S, B=(26^N)*2, E=11914220, we have a storage size of 1110 MB. The best case for storage but worse for access time is when all data pieces are in a small number of buckets, in big lists. In this case for N=5, S=11881376, U=1, B=(26^N)/2, E=200610 we have a storage size of 23 MB. In an average case of N=5, S=11881376, U=7510766, B=26^N, E=2679046 we have 682 MB of storage space needed. The sample for the above examples was the complete range of 5-Grams for the 26 lowercase English characters which are 26^5 = 11881376.

Our proposed technique now, the one that we call Dimensional Mapping, has a standard storage space, depending only on the length of the N-Grams. The idea is to store only the weight of each N-Gram with the N-Gram itself being the pointer to where it is stored. That is achieved by creating an N-dimensional array of integers where each character of the N-Gram is used as an index. So, in order to access the weight of the 5-Gram 'fdsgh' in the table DM we would just read the value in cell DM['f']['d']['s']['g']['h']. A very simple mapping is used between the characters and an integers: after a very strict cleansing process where we convert all characters in lowercase and discard all characters but the 26 in the English alphabet, we are just subtracting the ASCII value of 'a'. Due to the serial nature of the characters that gives us an integer between 0 and 26 that we can use as an index. A more complex mapping can be used in order to include more characters or even punctuation that we now ignore.

The Dimensional Mapping has a standard storage size requirement, dependent only on the length of the N-Grams as we mentioned before. The size it occupies can be estimated by the following formula:

$$(26^N) * 4 + \sum_{i=1}^{i=N} \left((26^{N-i}) * 12\right)$$

Equation 2: Dimensional Mapping size estimation with N being the length of N-Grams.

This may seem large but for the 5-Grams the estimated size is just 51 MB. Compared to the worst case of Hash Lists (1110 MB) or even the average case (682 MB) it seems like a huge improvement. This is caused due to the fact that the multidimensional array stores primitive values and not objects, which reduces the overhead greatly. Moreover, we can now say that accessing and updating a certain data piece can be done in O(1) time with absolute certainty, with no dependency on the data itself or a hash function. This had significant results in speeding up the execution times of the analysis, enabling us to look into streaming data and semi-supervised machine learning algorithms.

### 4.4 Results

We measured three main KPIs for the result comparison. Two of them (success ratio, kappa variable) were measuring the success ratio of classification and one (execution time) the algorithmic improvement. We present them bellow.

We run experiments on 5-Grams stored in classic ArrayList format, in Hash Lists and in Dimensional Mapping. After storing the N-Grams in these formats we applied a 10-fold cross validation on each one of the seven machine learning algorithms we chose: Naïve Bayesian Networks, C4.5, Support Vector Machines, Logistic Regression, Multilayer Perceptrons, Best-First Trees and Functional Trees. Then we recorded the three KPIs for each one of these 21 experiments. The results for the first two KPIs are shown in the bar chart that follows. In the same chart we have included the KPIs for a threshold based classification, using an arbitrarily set threshold.

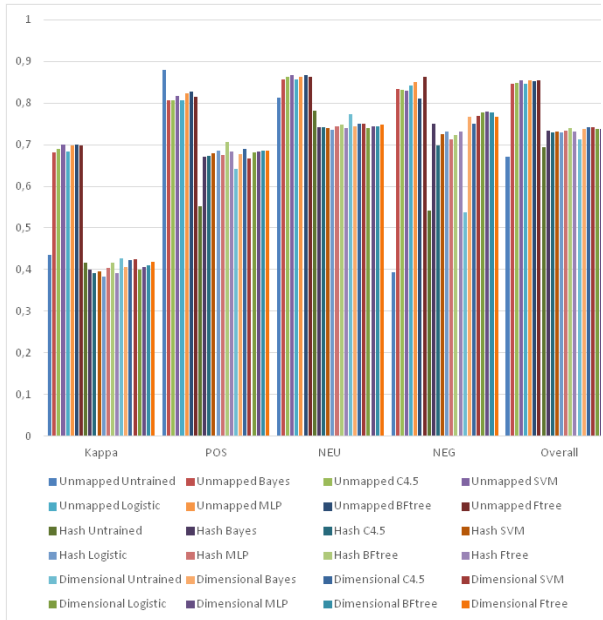

Figure 3: A comparison of the three KPIs as shown in the sentiment analysis experiments

As of the execution times the following table contains a summary of the results:

Table 1: Execution time in seconds summary - comparing for the various algorithms and techniques

|  | **ArrayLists** | **Hash List** | **Dimensional Mapping** |
|---|---|---|---|
| **Thresholds** | 1691 | 5 | 4 |
| **Naïve Bayes** | 12302 | 7 | 7 |
| **C4.5** | 21535 | 9 | 8 |
| **SVM** | 20662 | 147 | 177 |
| **Logistic Regression** | 22251 | 9 | 11 |
| **MLP** | 21224 | 41 | 48 |
| **BFTree** | 23319 | 25 | 19 |
| **FTree** | 22539 | 16 | 16 |

## 5. CONCLUSIONS

RADICAL platform, as presented in the current work, successfully combines citizens' posts retrieved through smartphone applications and Social Networks in the context of smart city applications, to produce a testbed for applying multiple analysis functionalities and techniques. The exploitation of resulting big aggregated datasets pose multiple challenges, with timely-efficient analysis being the most important. Focusing on data storage and representation, multiple techniques were examined in the experiments performed, in order to come up with the optimal algorithmic approach of Dimensional Mapping. In the future the authors plan to use even larger and more complex datasets, further leveraging on the effectiveness of these social networking services.

## ACKNOWLEDGEMENTS

This work has been supported by RADICAL and Consensus projects and has been funded by the European Commission's Competitiveness and Innovation Framework Programme under grant agreements no 325138 and 611688 respectively.